\documentstyle[epsfig,aps,12pt]{revtex}

\begin{document}

\title{One and Two-photon Physics with Relativistic Heavy Ions}
\author{C.A. Bertulani}
\address{Department of Physics, 
Brookhaven National Laboratory, Upton, NY 11973-5000, USA }
\date{\today }
\maketitle

\begin{abstract}
The physics of peripheral collisions with relativistic 
heavy ions (PCRHI) is reviewed. One- and two-photon processes are discussed.
\end{abstract}
\bigskip

   
\section{Peripheral Collisions}

In peripheral collisions of
relativistic heavy ions (PCRHI) \cite{BB94} the number of equivalent photons with 
energy $\omega $, $n(\omega )$,   has been calculated classically \cite{Fe24}, and
quantum-mechanically \cite{BB85}. 
It has been shown \cite{BB85} that for the electric dipole multipolarity, E1, the equivalent
photon number, $n_{E1}(\omega)$, coincides with the one deduced classically \cite{Fe24}
for very forward scattering. $E2$-photons are more abundant at low energies. But,  
in the extreme relativistic collisions the equivalent photon numbers for all
multipolarities agree, i.e, $n_{E1}(\omega )\sim n_{E2}(\omega )\sim
n_{M1}(\omega )\sim ...$. The cross sections for one- and two-photon
processes depicted in figure 1 are given approximately by
\begin{equation}
\sigma _{X}=\int d\omega \frac{n\left( \omega \right) }{\omega }\sigma
_{X}^{\gamma }\left( \omega \right) \;,\;{\rm and}\;\;\sigma _{X}=\int
d\omega _{1}d\omega _{2}\frac{n\left( \omega _{1}\right) }{\omega _{1}}\frac{%
n\left( \omega _{2}\right) }{\omega _{2}}\sigma _{X}^{\gamma \gamma }\left(
\omega _{1},\omega _{2}\right) \;,  \label{epa}
\end{equation}
where $\sigma _{X}^{\gamma }\left( \omega \right) $ is the photon-induced
cross section for the energy $\omega $, and $\sigma _{X}^{\gamma \gamma
}\left( \omega _{1},\omega _{2}\right) $ is the two-photon cross
section. 
For one-photon processes, e.g., Coulomb fragmentation, $\sigma
_{X}^{\gamma }\left( \omega \right)$ is localized in a small energy interval 
and one gets a cross section in the form $\sigma =A\ln \gamma
_{c}+B$, where $A$ and $B$ are coefficients depending on the system. The $ln \gamma_c$ factor
is due to the equivalent photon number (EPA), $n\left( \omega \right)$, which is approximately
independent of $\omega$ in the integral range of interest. As for the two-photon processes, 
besides the $ln^2 \gamma_c$ from $n_1$ and $n_2$, a third $ln \gamma_c$ arises from the 
integral over $\omega_1$ ($\omega_1$ and $\omega_2$ are related by energy conservation). 
Note that here we used $\gamma_{c} $ of a HI-collider, so that 
$\gamma =2\gamma _{c}^{2}-1$, with $\gamma_{c} $ the collider Lorentz gamma factor 
($\gamma _{c}\sim 100$ for RHIC/BNL).

\begin{center}
\begin{figure}[tbh]
{\epsfxsize=10cm \epsfbox{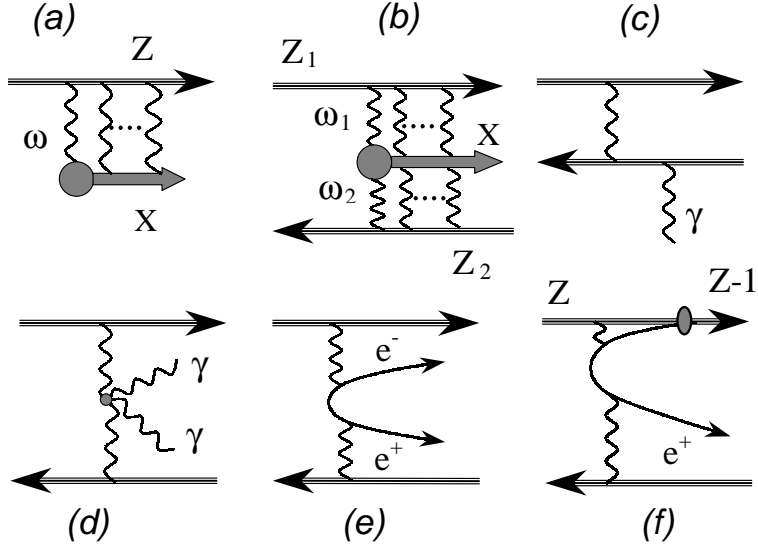}}
\caption{ PCRHI processes: (a) one-photon, (b) two-photon, (c) Bremsstrahlung,
(d) Delbr\"uck scattering, (e) pair-production, and (f) pair-production with
capture.}
\label{fig1}
\end{figure}
\end{center}

\section{Bremsstrahlung and Delbr\"{u}ck scattering}

Bremsstrahlung (fig. 1c) is a minor effect in PCRHI \cite{BB88}. The cross
section is proportional to the inverse of the square mass of the ions. Most
photons have very low energies. For 10 MeV photons
the central collisions (CCRHI) deliver 10$^{6}$ more photons than the PCRHI \cite{BB89}.
For a collider the Bremsstrahlung differential 
cross section is given
by $d\sigma _{\gamma }/d\omega =16Z^{6}\alpha ^{3}\left( 3\omega
A^{2}m_{N}^{2}\right) ^{-1}\ln \left( \gamma /\omega R\right) $, where $%
m_{N} $ is the nucleon mass, $\gamma =2\gamma _{c}^{2}-1$, where $\gamma
_{c} $ is the collider Lorentz gamma factor ($\gamma _{c}\sim 100$ for
RHIC/BNL), and $R$ is the nuclear dimension ($R\sim 2.4\times A^{1/3}$ fm) \cite
{BB89}.

For very low energy photons ($\omega \sim 100$ eV) the whole set of particles
in a bunch act coherently and a great number of Bremsstrahlung photons are produced.
This has been proposed as a tool for monitoring the bunch
dimensions in colliders \cite{Gin92}.

Delbruck scattering ($\gamma ^{\ast }+\gamma ^{\ast }\longrightarrow \gamma
+\gamma $) involves an additional $\alpha ^{2}$ as compared to pair
production and has never been possible to study experimentally. \ The cross
section is about 50 b for the LHC \cite{BB89} and the process is dominated
by high-energy photons, $E_{\gamma }\gg m_{e}$. A study of this process in
PCRHI is thus promising if the severe background problems arising from CCRHI
can be eliminated. No experiments of Bremsstrahlung or
Delbr\"{u}ck scattering in PCRHI have been performed so far. The total cross section
for Delbr\"{u}ck scattering ($\omega \gg m_{e}$) in colliders is given by $%
\sigma =2.54Z^{4}\alpha ^{4}r_{e}^{2}\ln ^{3}\left(
\gamma /m_{e}R\right) $, where $r_{e}=e^{2}/m_{e}$ is the classical
electron radius \cite{BB89}.

\section{Atomic ionization}

The cross sections for atomic ionization are very large, of order of
kilobarns, increasing slowly with the logarithm of the RHI energy. For a
fixed target experiment using naked projectiles one gets \cite{BB88}: $\sigma
_{I}=Z_{P}^{2}r_{e}^{2}\left( Z_{T}\alpha \right) ^{-2}\left[ 1.8\pi +9.8\ln
\left( 2\gamma /Z_{T}\alpha\right) \right] $, which decreases with the target
charge $Z_{T}$. This is due to the increase of the binding energy of $K$%
-electrons with the atomic charge. \ The first term is
due to close collisions assuming elastic scattering of the electron off the
projectile, while the second part is for distant collisions, with
impact parameter larger than the Bohr
radius. 
Recently, Baltz \cite{Bal01} has shown that the numerical factors in the 
equation above should be replaced by  $1.8 \rightarrow 1.74-1.83$ and $9.8 
\rightarrow 7.21$, respectively, when one includes
higher order terms in the perturbation series.
The probability to eject a $K$-electron is much larger than for
other atomic orbitals. \ Recent experiments have reported ionization cross
sections for $Pb^{81+}$ (33 TeV) beams on several targets \cite{Kr98}. In
this case, the role of projectile and target are exchanged in the previous
equation. In figure 2 we show the
results of this equation (dashed line) compared to 
the experimental data. Since the targets are
screened by their electrons, the discrepancy is expected. Even the most
detailed calculations by Anholt and Becker \cite{AB87} (solid line) or of
Baltz \cite{Bal01} yield
larger cross sections than the experimental data. 

\begin{figure}[thbp]
\begin{minipage}[t]{5.5cm}
{\epsfxsize=5.5cm \epsfbox{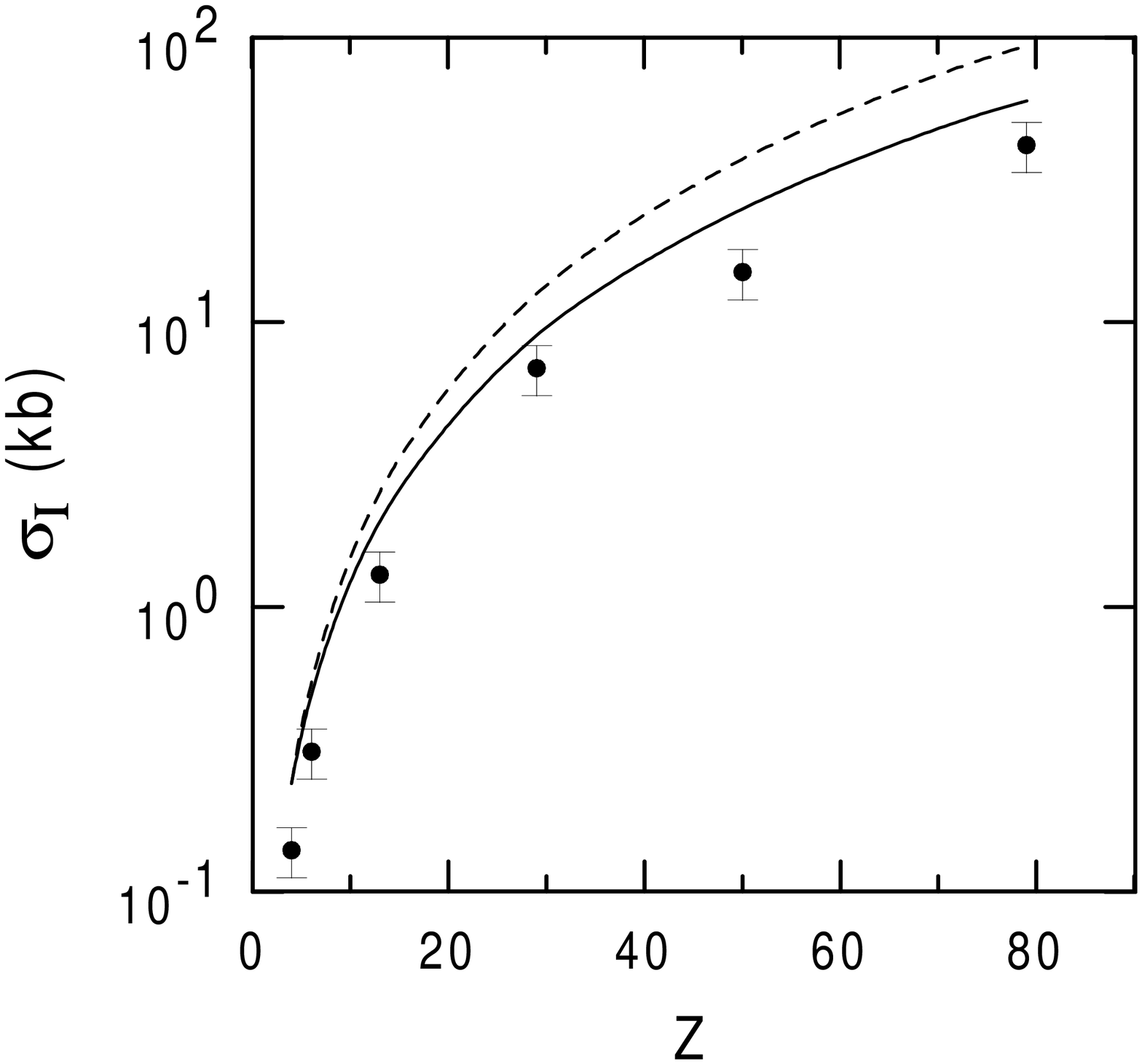}}
\caption[]{\small Atomic ionization cross sections 
for $Pb^{81+}$ (33 TeV) beams on several targets \cite{Kr98}.}
\label{fig2}
\end{minipage} \hfill
\vspace*{-68mm} \hspace*{5.5cm} \hfill
\begin{minipage}[t]{5.5cm}
{\epsfxsize=5.5cm \epsfbox{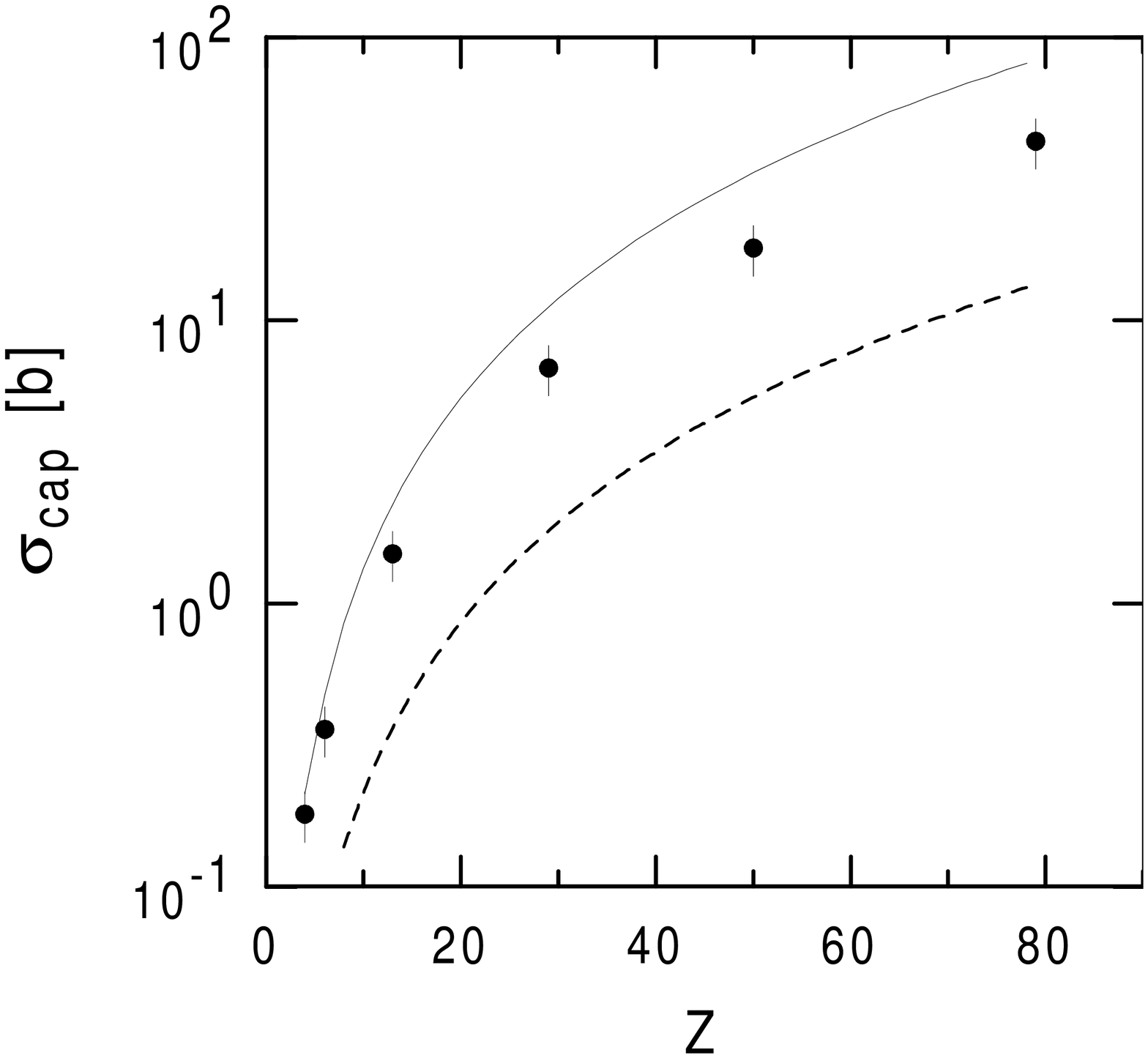}}
\caption[]{\small Pair production with capture 
for $Pb^{82+}$ (33 TeV) beams on several targets \cite{Kr98}.}
\label{fig3}
\end{minipage}
\end{figure}

Non-perturbative calculations, solving the time
dependent Dirac equation exactly, were first performed by Giessen and Oak
Ridge groups \cite{Bec83,BS85}. The main problem is to adequately treat 
the several channels competing with the ionization process,  
specially for atoms with more than one electron. Also, 
the effects of screening (static and dynamical) are hard to calculate.
On the experimental side, there are little data  available for a 
meaningful comparison with theory.

\section{Free and bound-free electron-positron pair production}

It has been demonstrated long
time ago   \cite{Fu33} that
to leading order in $ln \gamma $, the $e^+e^-$-pair production in PCRHI is given by
$\sigma =(28/27\pi)Z_{P}^{2}Z_{T}^{2}r_{e}^{2}\ln ^{3}\left( \gamma
/4\right) $.  A renewed interest in this process appeared with the
construction of relativistic heavy ion accelerators. For heavy ions with
very large charge (e.g, lead, or uranium) the pair production probabilities and
cross sections are very large. They cannot be treated to first order in
perturbation theory \cite{BB88}, and are also difficult to calculate. This resulted
in a great amount of theoretical studies \cite{RB91}. 

Replacing the 
Lorentz compressed electromagnetic
fields by delta functions, and working with light cone variables, one has
developed more elaborate calculations \cite{Ba97}, recently.  The latest
debate around the subject is the proper treatment of Coulomb distortion
of the lepton wavefunctions, and of production of n-pairs \cite{Ba97}.

An important phenomenon occurs when
the electron is captured in an atomic orbit of the projectile, or of the target.
In a collider this leads to beam losses each time a charge modified
nucleus passes by a magnet downstream \cite{BB89}. A striking
application of this process was the recent production of antihydrogen atoms
using relativistic antiproton beams \cite{Ba96}. Here the positron
is produced and captured in an orbit of the antiproton. Early calculations
for this process
used perturbation theory \cite{AB87,Bec87,BB88}. 
Some authors use non-perturbative approaches, e.g.,
coupled-channels calculations \cite{RB91}. Initially some discrepancy with perturbative
calculations were found, but later it was shown that non-perturbative
calculations agree with the perturbative ones at the  1\% level (see, e.g., first
reference of \cite{Ba97}).

The expression $\sigma =3.3\pi Z^{8}\alpha ^{6}r_{e}^{2}\left[
\exp \left( 2\pi Z\alpha \right) -1\right] ^{-1}\left[ \ln \left(
0.681\gamma/ {2}\right) -5/3\right] $ for pair production with electron
capture in PCRHI was obtained in ref. \cite{BB88}. The term $\left[ {...}\right]
^{-1}$ is the main effect of the distortion of the positron wavefunction.
It arises through the normalization of the continuum wavefunctions which
accounts for the reduction of the magnitude of the positron wavefunction
near the nucleus where the electron is localized (bound). Thus, the greater
the $Z$, the less these wavefunctions overlap. The above equation
predicts a dependence of the cross section in the form $\sigma =A\ln \gamma
+B$, where $A$ and $B$ are coefficients depending on the system. 
This dependence was used in the analysis of the
experiment in ref. \cite{Kr98}. In recent calculations, attention was given
to the correct treatment of the distortion effects in the positron
wavefunction \cite{He00}. In figure 3 we show the recent experimental
data of ref. \cite{Kr98} compared to the above equation and recent
calculations (second reference of \cite{He00}). These calculations also
predict a $\ln \gamma$ dependence but give larger cross sections
than in ref.\ \cite{BB88}. The comparison with the experimental data is not
fair since atomic screening was not taken into account. When screening is
present the cross sections will always be smaller up to a factor of 2 
\cite{BB88}. The conclusion here is that pair production with electron
capture is a process which is well treated in first order perturbation
theory. Again, the main concern is the correct treatment of distortion effects
(multiphoton scattering) \cite{He00}.

The production of para-positronium in heavy ion colliders
was calculated \cite{serbo}. The cross section at RHIC is about 18 mb. 
This process is of interest due to the unusual large transparency of the parapositronium
in thin metal layers.

\section{Relativistic Coulomb excitation and fragmentation}

Relativistic Coulomb excitation is becoming a popular tool for the
investigation of the intrinsic nuclear dynamics and structure of the
colliding nuclei, specially important in reactions
involving radioactive nuclear beams  \cite{Au98,BCH93}. 
Coulomb excitation and dissociation of such nuclei are 
common experiments in this field \cite{BB94,AB95}. The advantage is
that the Coulomb interaction is very well known. The real
situation is  more complicated since the contribution of the
nuclear-induced processes cannot be entirely separated in the experimental
data.  
The treatment of the dissociation problem by
nuclear forces is very model dependent, based
on eikonal or multiple Glauber scattering approaches \cite{BCH93,BBK92}. Among
the uncertainties are the in-medium nucleon-nucleon cross sections at
high-energies, the truncation of the multiple scattering process and the
separation of stripping from elastic dissociation of the nuclei \cite{HRB91}.
Nonetheless, specially for the very weakly-bound nuclei, relativistic
Coulomb excitation has lead to very exciting new results \cite{BCH93,BBK92}.

\begin{figure}[thbp]
\begin{minipage}[t]{5.5cm}
{\epsfxsize=5.5cm \epsfbox{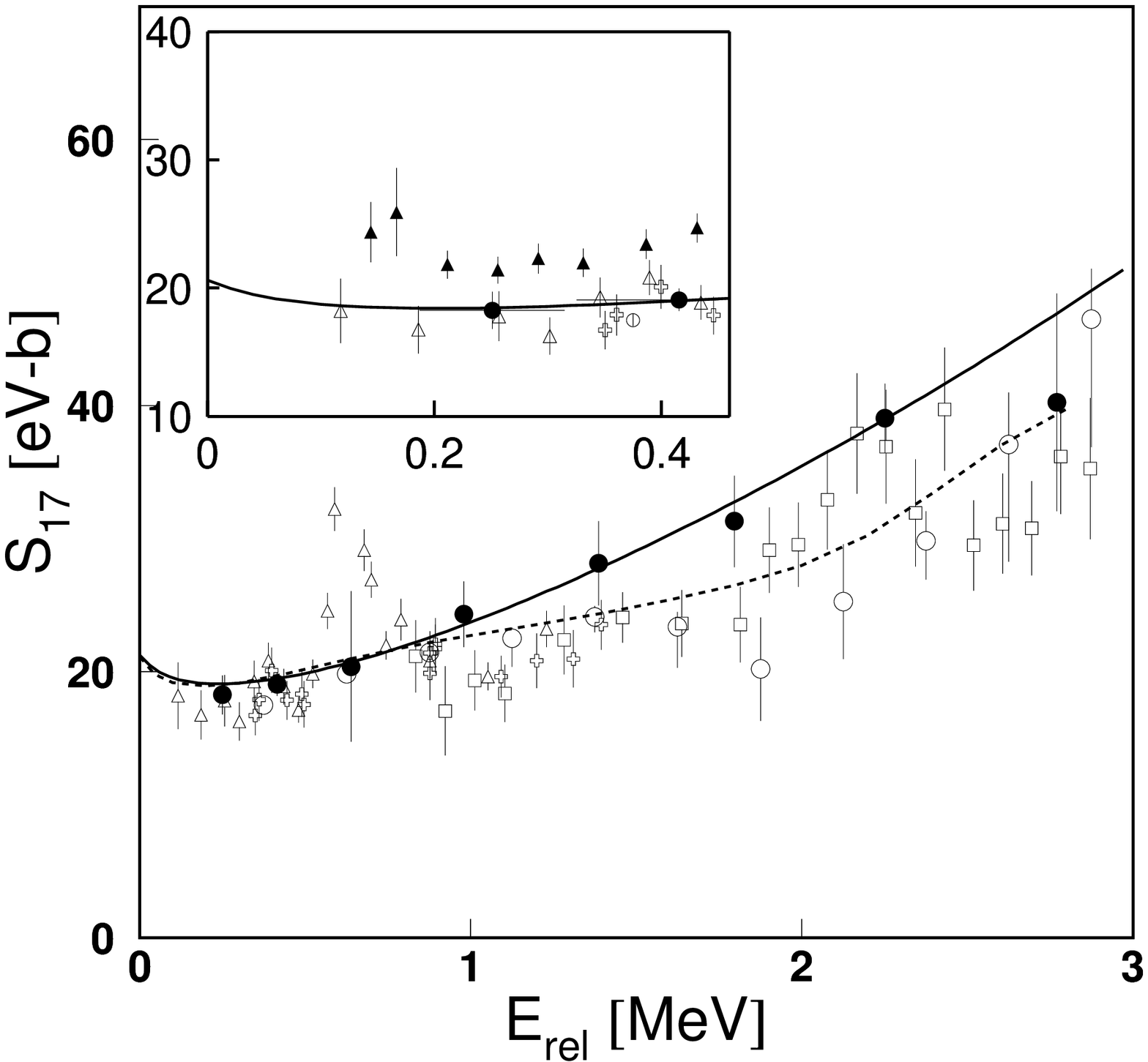}}
\caption[]{\small S-factors for the $^7Be(p,\gamma)^8B$ reaction.}
\label{fig4}
\end{minipage} \hfill
\vspace*{-68mm} \hspace*{5.5cm} \hfill
\begin{minipage}[t]{5.5cm}
{\epsfxsize=5.5cm \epsfbox{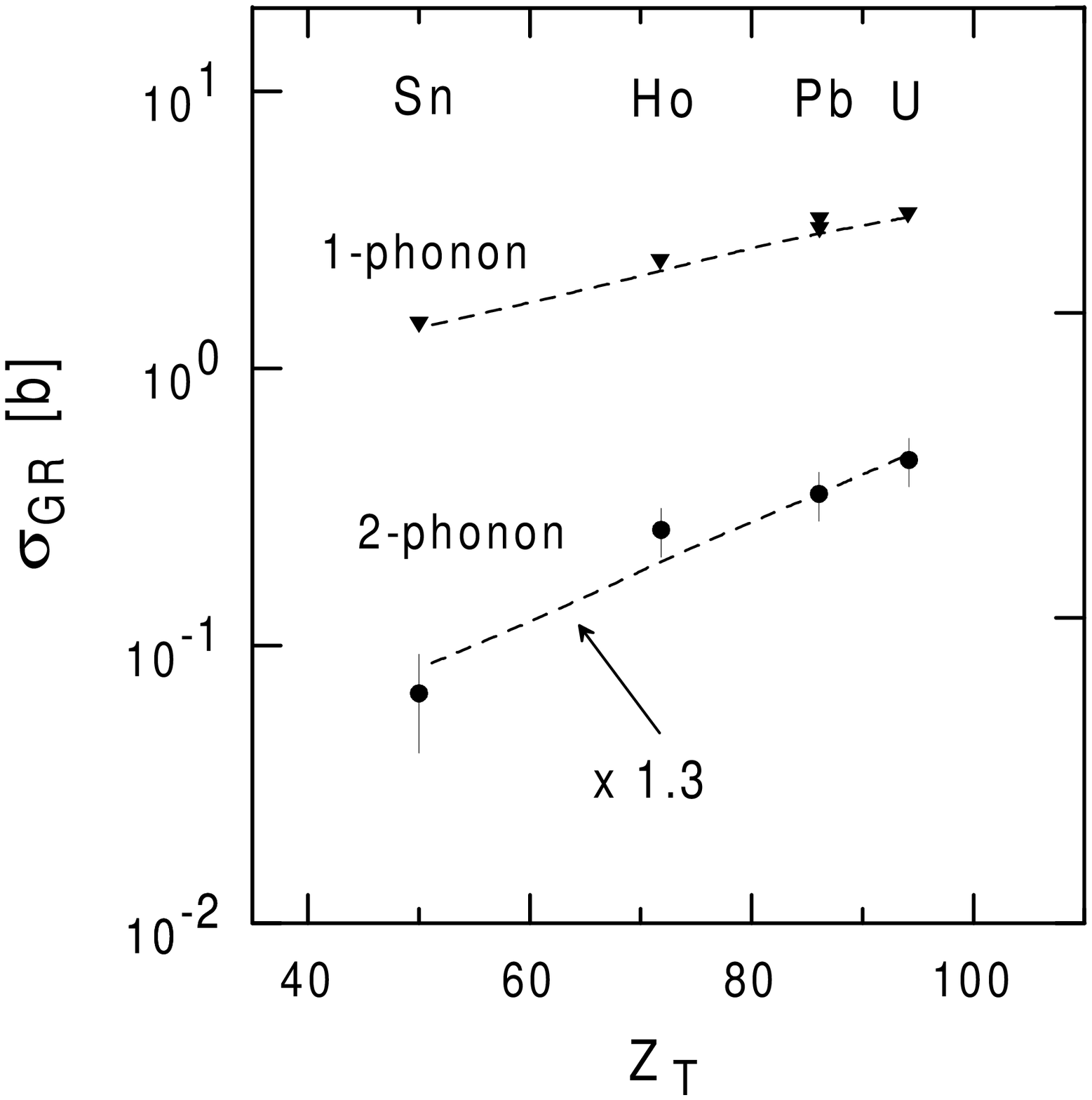}}
\caption[]{\small Cross sections for the excitation of the GDR and
the DGDR.}
\label{fig5}
\end{minipage}
\end{figure}

In the Coulomb breakup of weakly-bound nuclei one speculates if
the reaction proceeds via a single or multiple photon-exchange between the
projectile and the target. In the first case, perturbation theory gives a
direct relation between the data and the matrix element for
electromagnetic dissociation. Such matrix elements are the clearest probes
one can get about the nuclear structure of these nuclei. In the second case,
often called by post-acceleration effects \cite{BBK92}, one has to perform a
non-perturbative treatment of the reaction what complicates the extraction
of the electromagnetic (mainly E1) matrix elements. 
One expects to learn if
the Coulomb-induced breakup proceeds via a resonance or by the direct
dissociation into continuum states  \cite{BBK92}. There is a strong ongoing effort
to use the relativistic Coulomb excitation technique also for studying the
excitation of bound excited states in exotic nuclei, to obtain 
information on gamma-decay widths, angular momentum,
parity, and other properties of hitherto unknown states \cite{AB95}.

Radiative capture reactions are known to play a
major role in astrophysical sites, e.g., in a pre-supernova \cite{Rolfs}.
Some of these reactions, like for example, $^{7}Be\left( p,\gamma \right)
^{8}B$, can be studied via the inverse photo-dissociation reaction $%
^{8}B\left( \gamma ,p\right) ^{7}Be$. 
One often uses the astrophysical S-factor, defined by $%
S(E)=E\sigma \left( E\right) \exp \left[ -2\pi \eta \left( E\right) \right] $, 
where $\eta \left( E\right) =Z_{1}Z_{2}e^{2}/\hbar \sqrt{2\mu _{12}E}$,
where $E$ is the relative kinetic energy of the two nuclei.
The matrix
elements involved in the dissociation processes are the same as those
involved in the absorption by real photons \cite{BB88}. 
One of the
experiments using this technique was performed at the GSI/Darmstadt \cite
{Iw99}. The S-factor obtained in this experiment is shown in figure 4 as
solid circles. 

A giant dipole resonance (GDR) occurs in nuclei at
energies of 10-20 MeV. Assuming that these are harmonic vibrations of 
protons against neutrons, one expects that DGDRs (Double Giant Dipole Resonances), 
i.e., two giant dipole
vibrations superimposed in one nucleus, will have exactly
twice the energy of the GDR \cite{BB88,Au98}. 
A series of
experiments at the GSI/Darmstadt has obtained energy spectra, cross
sections, and angular distribution of fragments following the decay of the
DGDR. 
It was shown that the experimental cross sections are 
about 30\% bigger than
the theoretical ones. This is shown in figure 5 where the cross sections
for the excitation of 1-phonon (GDR), $\sigma _{1}\sim 2\pi S\ln \left[
2\gamma A_{T}^{1/3}\left( A_{P}^{1/3}+A_{T}^{1/3}\right) ^{-1}\right] $,
while for the 2-phonon state it is $\sigma \sim S^{2}\left(
A_{P}^{1/3}+A_{T}^{1/3}\right) ^{-2}$, where $S=5.45\times
10^{-4}Z_{P}^{2}Z_{T}N_{T}A_{T}^{-2/3}$ mb. The dashed lines of figure 5
are the result of more elaborate calculations \cite{Au98}. The GSI
experiments are very promising for the studies of the nuclear response in
very collective states. One should notice that after many years of study
of the GDRs and other collective modes, the width of these states are still
poorly explained theoretically, even with the best microscopic approaches
known sofar. The extension of these approaches to the study of the width of
the DGDRs will be helpful to improve such models. 

In colliders it was shown that the mutual Coulomb excitation of the ions
(leading to their simultaneous fragmentation) is a powerful tool for beam
monitoring \cite{seb1}. A recent measurement at RHIC by Sebastian White and
collaborators \cite{seb2}, using the Zero Degree Calorimeter to measure
the neutron decay of the reaction products, has proved the
feasibility of the method. The theoretical prediction of about 3 b for this process, 
agrees quite well with the experimental results.    

The DGDR contributes only to about 10\% of the total fragmentation cross
section induced by Coulomb excitation with relativistic heavy ions. The main
contribution arises from the excitation of a single GDR, which decays mostly
by neutron emission. This is also a potential source of beam loss in relativistic
heavy ion colliders \cite{BB94}, and an important fragmentation mode of 
relativistic nuclei in cosmic rays.

\section{Meson and hadron production}

The production of heavy lepton pairs ($\mu^+\mu^-$, or $\tau^+\tau^-$), or of
meson pairs (e.g., $\pi^+\pi^-$) can be calculated using the second of equation 
(1). One just needs the cross sections for $\gamma\gamma$ production of these pairs.
Since they depend on the
inverse of the square of the particle mass \cite{BB89},
the pair-production cross sections are much smaller in this case. 
The same applies to single meson production by $\gamma\gamma$ fusion. The
$\gamma\gamma$ cross section is given by $\sigma_{\gamma\gamma\rightarrow M}= 8 \pi^2
(2J+1)\Gamma_{M\rightarrow\gamma\gamma}\delta(W^2-M^2)/M$, where $J$, $M$, and 
$\Gamma_{M\rightarrow\gamma\gamma}$ are the spin, mass and two-photon decay width of the
meson, $W$ is the c.m. energy of the colliding photons \cite{BB89}. 
In ref. \cite{BB88} the following equation was obtained for the production of mesons
with mass $M$ in HI colliders: 
$\sigma =   Z^4\alpha^2 (128 \Gamma_{\gamma\gamma}/3
M^3) \ \ln^3 \left( 2\gamma \delta / M R \right) $, where $\delta=0.681...$
Later \cite{BF90} it is shown that a more detailed account of the space geometry
of the two-photon collision is necessary, specially for the heavier mesons.

A careful study of the production of meson pairs and single mesons in PCRHI was
performed recently in ref. \cite{RN00} (see also, \cite{spencer1}). 
In table I we show the magnitude of the cross sections for single meson production 
at RHIC and at LHC \cite{RN00}. Also shown are the cross sections due to difractive
processes (pomeron-pomeron exchange). We see that they are several orders of
magnitude smaller than those from $\gamma\gamma$ fusion. The cross sections for the
production of $\eta_c$, $\eta_c'$ and $\eta_b$ are very small due to their higher masses.
Similar studies have been done for meson production in $\gamma$-nucleus interactions.
Particles like $\Delta$, $\rho$, $\omega$, $\phi$, $J/\Psi$, etc, can be produced in this
way \cite{BHT98}.  
However, the one-photon exchange processes are more effective in the production of mesons
in PCRHI. One considers the interaction between the photon from one nucleus with a Pomeron
from another. These photon-Pomeron interactions were calculated in ref. \cite{spencer2}
and are shown in table II.

The possibility to produce a Higgs boson via $\gamma\gamma$ fusion was suggested in
ref. \cite{Gr89}. The cross sections for LHC are of order of  1 nanobarn, about the 
same as for
gluon-gluon fusion. But, the two-photon processes can also produce $b\bar{b}$ pairs which
create a large background for detecting the Higgs boson. A good review of these topics
was presented in ref. \cite{BHT98}.

The excitation of a hadron in the field of a nucleus is another useful tool to study
the properties of hadrons. It has been used for example to obtain the lifetime of the
$\Sigma_0$ particle by measuring the (M1) excitation cross section for the process 
$\gamma + \Lambda \rightarrow \Sigma_0$ \cite{Dy77}. The vertex 
$\gamma \rightarrow 3\pi$ has been
investigated  \cite{Ant87} in the reaction of pion pair production by pions in the 
nuclear Coulomb field: $\pi^- +Z \rightarrow \pi^- +\pi^0 +Z$. Also, the $\pi^-$ 
polarizability has been studied in the reaction $\pi^- +Z \rightarrow \pi^- +\gamma+Z$
\cite{Ant83}. Other unexplored possibilities includes the excitation of a nucleon to
a $\Delta$-particle in the field of a heavy nucleus in order to disentangle the $M1$
and $E2$ parts of the excitation. 

As for meson-production in PCRHI there are several planned experiments at RHIC (see
the contribution of Spencer Klein to these proceedings), as well as for the future
LHC \cite{spencer3}. These machines were designed for study hadronic processes. But, as 
have shown in this brief review, they can also be used for studying very interesting
phenomena induced in peripheral collisions.

\section*{Acknowledgment(s)}
The author is a fellow of the John Simon Guggenheim Memorial Foundation.
This work has been authored under Contract N0. DE-AC02-98CH10886 
with the DOE/USA and  contract No. 41.96.0886.00 with the
PRONEX/Brazil.

\begin{table}[thb]
\caption{ Two-photon meson production at RHIC and at LHC. Masses are in MeV,
decay widths in keV, and cross sections in mb. The cross sections are for
$\gamma\gamma$ and pomeron-pomeron (${\cal PP}$) exchange processes, respectively.}
\par
\begin{tabular}{lllllll}
\hline
&  &  &  &  \\
Meson & M & $\Gamma_{X\rightarrow\gamma\gamma}$ & $RHIC_{\gamma\gamma}$ & $%
LHC_{\gamma\gamma}$ & $RHIC_{\cal PP}$ & $LHC_{\cal PP}$\\ \hline
&  &  &  &  \\[-10pt] 
$\pi^0$ & 135 & $8 \times 10^{-3}$ & 7.1 & 40 & 0.05 & 0.367 \\ 
$\eta$ &  547 & 0.463 & 1.5 & 17 & 0.038 & 0.355 \\ 
$\eta'$ & 958 & 4.3 & 1.1 & 22 & 0.04 & 0.405 \\  \hline
\end{tabular}
\end{table}

\begin{table}[thb]
\caption{ One-photon vector meson production  cross sections in mb at RHIC and at LHC. }
\par
\begin{tabular}{lllll}
\hline
&  &  &  &  \\ 
Meson & RHIC & LHC  \\ \hline
&  &    \\[-10pt] 
$\rho^0$ & 590 & 5200  \\ 
$\omega$ &  59 & 490 \\ 
$\phi$ & 39 & 460 \\  
$J/\Psi$ & 0.29 & 32 \\  
\hline
\end{tabular}
\end{table}

\end{document}